\PassOptionsToPackage{english}{babel}
\documentclass[]{beilstein}
\usepackage{graphicx,amsmath,color}
\usepackage{xspace}
\usepackage{amssymb}
\usepackage{sidecap}
\usepackage{hyperref}
\usepackage{epstopdf}

\begin{document}
\title{Dynamics and fragmentation mechanism of {\Pt} on SiO$_2$ Surfaces}
\author{Kaliappan Muthukumar}

\affiliation{Institut f\"ur Theoretische Physik, Goethe-Universit\"at, Max-von-Laue-Stra{\ss}e 1, 60438 Frankfurt am Main, Germany}

\author{Harald O. Jeschke}

\affiliation{Research Institute for Interdisciplinary Science, Okayama University, 3-1-1 Tsushima-naka, Kita-ku, Okayama 700-8530, Japan}

\author*[1]{Roser Valent\'\i}{valenti@itp.uni-frankfurt.de}

\date{\today}
\newcommand{\wc}{W(CO)$_6$}
\newcommand{\mc}{M(CO)$_6$}
\newcommand{\bc}{${\beta}${-}cristobalite}
\newcommand{\methyl}{CH$_{3}$}
\newcommand{\sio}{SiO$_2$}
\newcommand{\FOH}{FOH - SiO$_2$ }
\newcommand{\POH}{partially hydroxylated SiO$_2$}
\newcommand{\Pt}{(CH$_{3}$-C$_{5}$H$_{4}$)Pt(CH$_{3}$)$_{3}$}
\newcommand{\cp}{methylcyclopentadienyl}
\newcommand{\Cp}{cyclopentadienyl}
\newcommand{\Me}{methyl}
\newcommand{\otwo}{O$_2$ }

\maketitle
\justify

\begin{abstract}
  The interaction of {\Pt}
  ((methylcyclopentadienyl)trimethylplatinum)) molecules on fully
  and partially hydroxylated {\sio} surfaces, as well as the dynamics of this interaction were investigated using density functional theory (DFT) and finite temperature
 DFT-based molecular dynamics simulations. Fully and partially hydroxylated surfaces represent substrates before and after electron beam treatment and this study
  examines the role of electron beam pretreatment on the substrates
  in the initial stages of precursor dissociation and formation of Pt deposits.
  Our simulations show that on fully hydroxylated
  surfaces or untreated surfaces, the precursor molecules remain
  inactivated while we observe fragmentation of {\Pt} on partially
  hydroxylated surfaces.  The behavior of precursor molecules on the
  partially hydroxylated surfaces has been found to depend on the
  initial orientation of the molecule and the distribution of surface
  active sites.  Based on the observations from the simulations and
  available experiments, we discuss possible dissociation channels
  of the precursor.
 \end{abstract}

\keywords{ {\Pt}; FEBID; EBID; Deposition; Precursor; Dissociation}

\section{Introduction\label{Introduction}}

Nanoscale device applications require a growth of regular or specially
patterned transition metal nanodeposits.  Electron beam induced
deposition (EBID), is a size and shape selective deposition process
capable of writing low dimensional, sub-10 nm patterns on conducting
and insulating substrates with tunable electronic
properties~\cite{Wnuk2011,Utke2008,Wnuk2009,weber:461,Weirich2013}.
However, the deposits obtained often contain less than 50 atomic \% of
metal which is detrimental to their conductivity.  The incomplete
dissociation of the precursor molecules on the substrate during the
deposition process leaves a significant organic residue, thus
impairing the quality of the
deposits~\cite{incompletedissociation1,Wnuk2009}. This
lowers the range of applicability of EBID for nanotechnological
applications. Several postfabrication approaches (such as annealing,
post-deposition annealing in O$_{2}$, exposure to atomic hydrogen,
post deposition electron irradiation) were proposed as viable
techniques~\cite{Huthreview2012,Lewis2015}, but these approaches are
not completely free from reproducibility issues.  Therefore, to improve
the metal content and to address the nature of the organic contamination,
a fundamental understanding of how the molecules behave on the
substrates is necessary.  This will be helpful, either to modify the
existing class of precursor materials or to design a novel set of
precursors, specific for electron beam deposition.  To address this,
we have made a series of DFT studies in which
we considered fully and partially hydroxylated {\sio} surfaces as a
representative for untreated and electron beam pretreated substrates
and investigated the adsorption~\cite{Muthukumar2011,Muthukumar2012}
and dynamics of several carbonyl precursors~\cite{Muthukumar2014}.

{\Pt}, in which Pt bonds directly to three methyl groups and a
methylated {\Cp} ring is a widely used precursor to obtain Pt
deposits.  Although the dissociation mechanism of {\Pt}, leading to
the Pt deposit remains unknown, studies for a family of precursors
similar to {\Pt} in atomic layer deposition (ALD) conditions are
available~\cite{ald-smgeorge}. The studies in this review fairly agree
that 1) the presence of surface hydroxyl groups are the source for the
protons that help in the evolution of H$_{2}$, CH$_{4}$ and H$_{2}$O
during the deposition process and 2) the molecules dissociate or
associate through a ligand exchange process~\cite{mastail_mechanism}.
Recently, examination of the behavior of this molecule on fully hydroxylated {\sio}
surfaces has shown that the molecule remains
physisorbed~\cite{Juan2012}. These static $T=0$~K DFT computations
provide insights on the bonding of {\Pt} to this surface, but are
limited to the adsorption behavior of the molecule.  Several questions
remain, such as the behavior of {\Pt} on electron beam pretreated
substrates, the role of temperature in the dissociation of the
precursor on untreated surfaces, and the possible mechanism by which
the precursor dissociates on the electron beam pretreated surfaces.

Hence, in order to extend the knowledge on the adsorption and to address
the open questions in the deposition process, in this study we use DFT
and finite temperature DFT-based molecular dynamic simulations (MD)
and investigate the adsorption behavior of {\Pt} on fully and
partially hydroxylated {\sio} surfaces.  We focus on explaining the
initial reactions and the possible fragmentation pathways of the {\Pt} molecule 
on the {\sio} surface and we explain the nature of organic contamination 
in the deposits.

\section{Computational Details}\label{Formalism}

DFT calculations for the substrates, {\Pt} precursor molecules, and
the combined substrate/precursor molecule system were performed using
the projector augmented wave (PAW)
method~\cite{Bloechl1994,Kresse1999} as implemented in the Vienna
Ab-initio Simulation Package
(VASP-5.2.11)~\cite{Kresse1996a,Kresse1996b,Kresse1993}.  The
generalized gradient approximation in the parameterization of Perdew,
Burke and Ernzerhof~\cite{Perdew1996} was used as approximation for
the exchange and correlation functional.  In addition, dispersion
corrections~\cite{Grimme2006} were used to simulate the long range van
der Waals interactions.  All calculations were performed in the scalar
relativistic approximation.  A kinetic energy cut-off of 400~eV was
used and all ions were fully relaxed using a conjugate gradient scheme
until the forces were less than 0.01~eV/{\AA}. In the geometry
optimizations for the molecule and the surface models the Brillouin
zone was sampled at the ${\Gamma}$ point only.  Spin polarization was
considered for all calculations. Different spin states (i.e. in each
case the two lowest possible spin states) were considered, and only
the results of the ground state are reported below.  The adsorption
energy ($E_{\rm A}$) was defined as $E_{\rm A}\equiv{\Delta}E = E_{\rm
  total}- E_{\rm substrate}- E_{\rm adsorbate}$, where $E_{\rm
  total}$, $E_{\rm substrate}$, and $E_{\rm adsorbate}$ are the
energies of the combined system (adsorbate and the slab), of the slab,
and of the adsorbate molecule in the gas phase in a neutral state,
respectively.

The most stable systems were considered for studying the dynamics of
the adsorbed precursor molecule. MD simulations were performed for 20
ps on a canonical ensemble at a finite temperature of $T = 600~K$
using the Nose-Hoover thermostat~\cite{JCC:JCC21057}. The temperature
$600\,{\rm K}$~\cite{temperature} was chosen in accordance with a
typical experiment used for {\Pt} deposition.  The Verlet algorithm in
its velocity form with a time step of $\Delta t =1$~fs was used to
integrate the equations of motion. For these simulations, we have used
a reduced (2x2x2) {\sio} supercell so as to reduce the computational
time. For reaction modeling studies, all species
(transition states (TS) and intermediates (INT)) in the proposed
reaction paths in Fig.~\ref{fig:scheme1} were traced at PM6 (Parametrized Model 6) level
using the Berny algorithm implemented in Gaussian-09 software~\cite{go9}.

\section{Results and Discussion}
\subsection{Adsorption of {\Pt} on {\sio} substrates}
The interaction of the precursor molecule with the substrate, in
general, depends on both the orientation of the adsorbate and the
adsorption site on the substrate.  The interaction of {\Pt} with the
fully hydroxylated {\sio} substrate surfaces was investigated by
placing the molecule with different orientations on several bonding
sites and the most stable configuration is the one in which the {\cp} ring and
two of the methyl groups that are directly bonded to Pt are oriented
towards the substrate~\cite{Juan2012}.

\begin{figure*}
	\centering
	\includegraphics[width=0.96\textwidth]{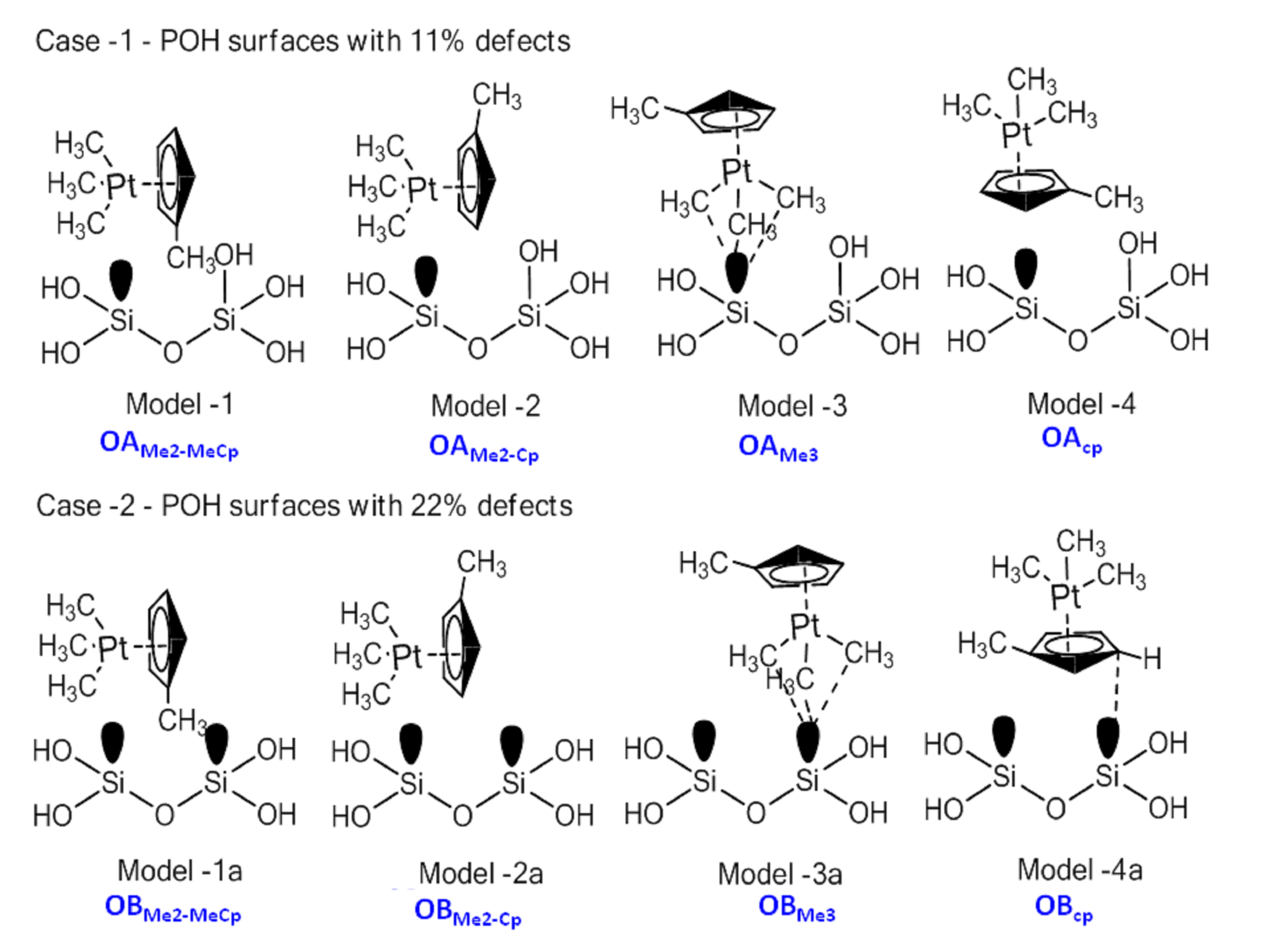}
	\caption{Schematic representation of the initial orientations
		of {\Pt} on 11\% (Model-1 to 4,  upper panel) and 22\%
		(Model-1a to 4a, lower panel) partially dehydroxylated
		surfaces.}
	\label{fig:configuration}
\end{figure*}

\begin{figure}
	\centering
	\includegraphics[width=0.9\textwidth]{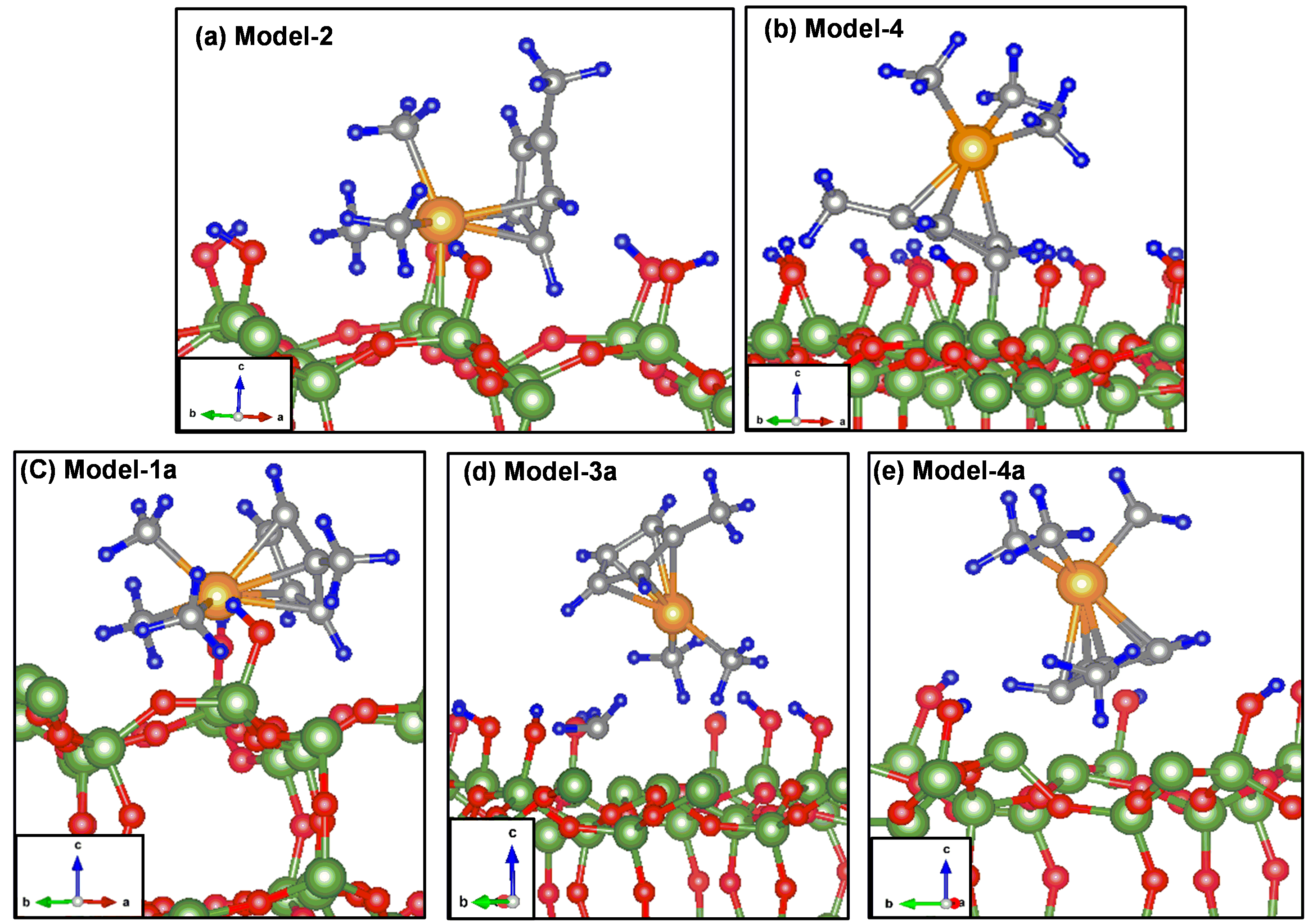}
	\caption{Relaxed structures of {\Pt} on the 11{\%} ((a), (b))
		and 22{\%} ((c)-(e)) partially dehydroxylated surfaces.  The
		labels correspond to the initial configurations shown
		in Fig.~\ref{fig:configuration}.  In the most stable
		configuration for the 22\% case (panel d) one of the three {\Me}
		groups bonded to Pt dissociates spontaneously during
		relaxation.  Color code: green - Si, red - O, orange - Pt,
		gray - C and blue - H throughout this manuscript.  To better
		display the changes occurring the system in 2D view, the
		snapshots have slightly different orientations in the $ab$
		plane (note the coordinate systems).}
	\label{fig:pohorientation}
\end{figure}

In a similar way, the bonding of {\Pt} precursor molecules on partially
hydroxylated {\sio} surfaces were investigated.  The initial
configurations of {\Pt} considered for simulations on the partially
hydroxylated surfaces are shown schematically in
Figure~\ref{fig:configuration}.  We simulate the precursor adsorption on
two different partially hydroxylated surface models, that differ in
the number of available active sites on the surfaces; 11\%
(Figure~\ref{fig:configuration} upper panel) and 22\%
(Figure~\ref{fig:configuration} lower panel).
On these surfaces, two sets of {\Pt} orientations were considered; (1)
reclining orientations (Model-1/1a and Model-2/2a)), that differ in
the orientation of the {\cp} ring of {\Pt}, and (2) upright
configurations (Model-3/3a) and Model-4/4a)) in which either three of
the {\methyl} groups bonded to Pt or the centroid of the {\cp} ring
bonds to the substrate.  These initial configurations are allowed to
relax without any constraints. The configurations where {\Pt} is
placed on the bridging sites move spontaneously to on-top sites
during geometry relaxation and hence are not discussed further.
The calculated adsorption energies ($E_{\rm A}$) are summarized in Table~\ref{T:Table1}
and the configurations in Fig.~\ref{fig:pohorientation}.

\begin{table}
\label{T:Table1}
\begin{center}
	\caption{Adsorption energies of {\Pt} on partially hydroxylated {\sio} surfaces with 11 and 22\% defects.  The cases listed correspond to configurations in Fig.~\ref{fig:configuration}. All values in eV/unit cell.}
\begin{tabular}{| c | c | c | }
\hline
\textbf{Cases} & \multicolumn{2}{ c |}{\textbf{\sio\ surfaces with}}  \\ 
\cline{2-3}
			& \textbf{11\% defects  } & \textbf{ 22\% defects} \\
			\hline
			Model-1/1a    &  -0.680  &  -2.321   \\ \hline
			Model-2/2a    &  -1.443  &  -2.297  \\ \hline
			Model-3/3a     &  -0.332  &  -2.769  \\ \hline
			Model-4/4a        &  -1.458  &  -2.360  \\ \hline
			
\end{tabular}
\end{center}
\end{table}

The calculated values of ($E_{\rm A}$) of {\Pt} on 11\% partially
dehydroxylated surfaces indicates that the most stable configurations
are Model-2 (OA$_{Me_{2}Cp}$ $E_A=-1.443$~eV) and Model-4 (OA$_{Cp}$, $E_A=-1.458$~eV) given
in Figure~\ref{fig:pohorientation}~(a) and (b)). The configurations are
equally stable within computational error. 
The configurations Model-1 (OA$_{Me_{2}MeCp}$, $E_A=-0.680$~eV) and Model-3 (OA$_{Me_{3}}$ , $E_A=-0.332$~eV) are less stable and are not considered for discussion. In Model-4, where the centroid of the {\cp} ring is oriented on top of a Si atom in the initial configuration, a spontaneous bond formation between one of the
carbon atoms of the ring and the surface Si atom is observed during
relaxation. There has been no evidence of any further dissociation.

Bonding of {\Pt} has also been considered on the 22\% partially
dehydroxylated surfaces and the relaxed configurations are
shown in Fig.~\ref{fig:pohorientation}~(c-e).
The calculated adsorption energies for the reclining configurations 
Model-1a (OB$_{Me_{2}MeCp}$, and Model-2a (OB$_{Me_{2}Cp}$) are -2.321 and -2.297 eV, respectively. We do not observe any spontaneous dissociation on these configurations .
However, when the relaxations are started with Model-3a orientation
(OB$_{Me_{3}}$) a stronger adsorption
with an $E_{\rm A}=-2.769$~eV is observed. In this case, one of the
three {\methyl} groups that bonds directly to the Pt atom dissociates,
as has been observed on the experimental investigations
~\cite{Wnuk2009,Egger,Liang,Koplitz,Pt2014}.  The detached {\methyl}
group was found to bind to one of the surface active sites.
However, this situation was not observed on 11\%
partially dehydroxylated surfaces indicating that the dissociation is
assisted by the neighboring active site, which is present on the 22\%
partially dehydroxylated surfaces. The next most stable configurations
on the 22\% partially dehydroxylated surface are obtained from
Model-4a ((OB$_{Cp}$) $E_{\rm A}=-2.360$~eV).

A comparison of the calculated adsorption energies for the 11\% and
22\% cases indicates that the molecules are more stable on the latter.
Furthermore, no dissociation on the 11\% dehydroxylated surfaces
was  observed.  These observations illustrate that the orientation of
{\Pt} on {\sio} surfaces, and the availability and location of active
sites on the surface are crucial factors in dictating the dissociation
of precursors and the growth mechanism of deposits.

\subsection{Dynamics of the {\Pt} Precursor on {\sio} substrates}

\begin{figure}
	\includegraphics[width=0.9\textwidth]{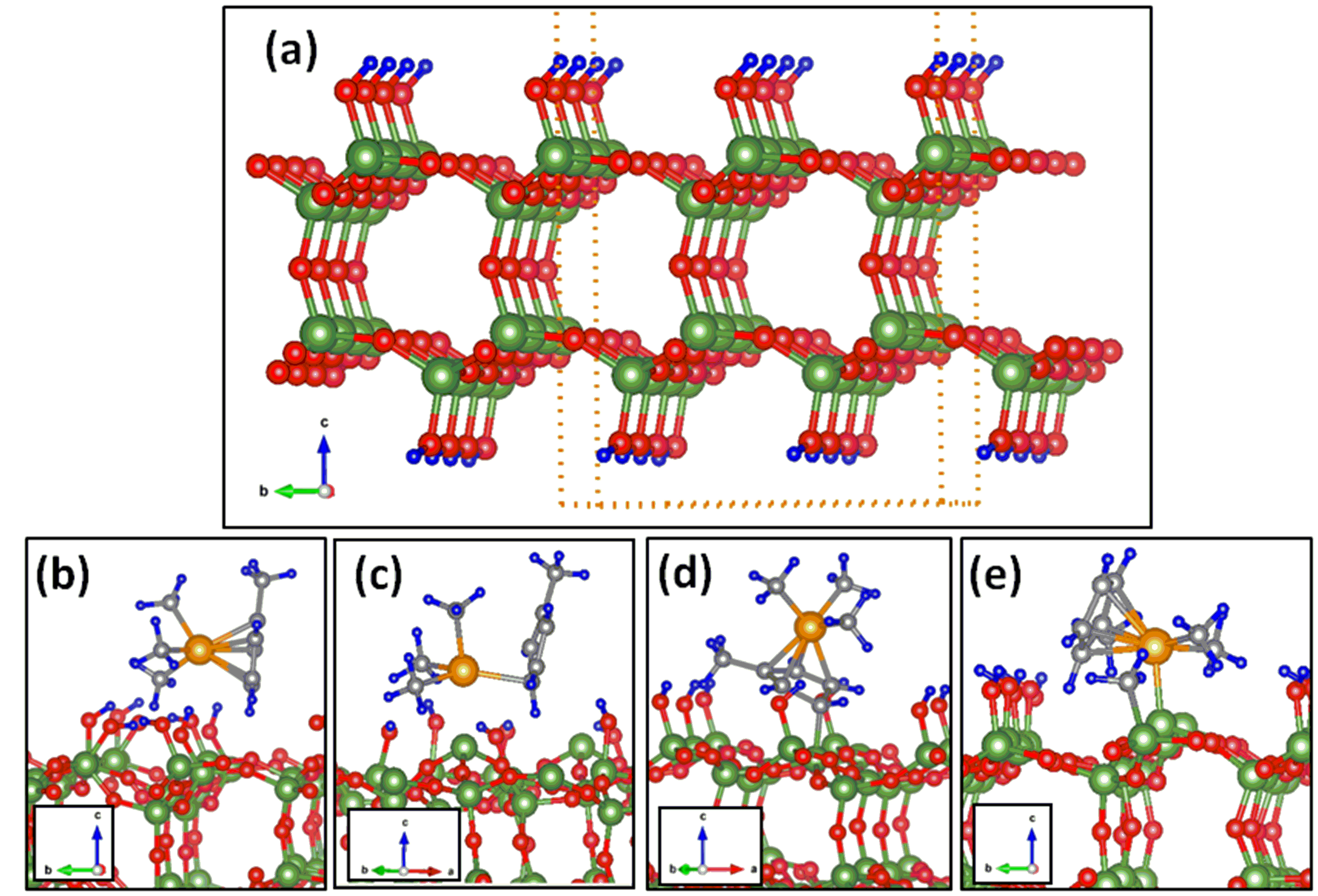}
	\caption {(a) Reduced supercell with the fully hydroxylated {\sio}
		considered for the molecular dynamics simulations. (b-e)
		Most stable configurations of {\Pt} (b) on the fully
		hydroxylated surface, (c) and (d) 25\% and (e) 50\% partially
		dehydroxylated surfaces.}
	\label{fig:2x2cellandorientation}
\end{figure}

To  gain further insight on the growth process of Pt deposits, dynamics of
{\Pt} on both fully and partially hydroxylated surfaces were
simulated. Due to the high computational expense however, a reduced
$2\times 2\times 2$ supercell of {\sio} (see
Figure~\ref{fig:2x2cellandorientation}~(a)), was used. This reduced
supercell provides a closer packing of precursor molecules (the
nearest neighbor distance between two Pt atoms in {\Pt} is ca. 8~{\AA}
in the reduced cell compared to ca. 15~{\AA} in the original supercell) and an
enhanced concentration of surface hydroxyl vacancies is provided in
the reduced cell on partially dehydroxylated surfaces (i.e, 25\% and
50\% compared to 11\% and 22\%).  However, the environment around
these sites is similar (i.e., the sites are isolated on the 25\%
partially dehydroxylated surface and located adjacent to each other on
the 50\% partially dehydroxylated surface).

\begin{table}
	\begin{center}
		\caption{Comparison of {\Pt} adsorption energies on the
			$3\times 3\times 4$ supercell and the reduced	$2\times 2\times 2$ supercell. For the fully hydroxylated surfaces, the values are taken from ref.~\cite{Juan2012}. Model-2 and 4 are the stable configurations of {\Pt} 
			on partially hydroxylated surfaces. All values in eV. }\label{tab:secondtable}
		\begin{tabular}{| c | c | c | }
			\hline
			\textbf{Cases} & \multicolumn{2}{ c |}{\textbf{Supercells}}  \\ 
			\cline{2-3}
			& \textbf{$3\times 3\times 4$ } & \textbf{ $2\times 2\times 2$} \\
			\hline
			Fully hydroxylated  $-$ (O$_{Me_{2}MeCp}$) &  -0.650  &  -1.02   \\ \hline
			Fully hydroxylated  $-$ (O$_{Me_{2}Cp}$)   &  -0.596  &  -0.845  \\ \hline
		    Model $-$  2                               &  -1.443  &  -1.494  \\ \hline
			Model $-$  4                               &  -1.458  &  -1.647  \\ \hline
					
		\end{tabular}
	\end{center}
\end{table}


\begin{figure}
	\includegraphics[width=1.0\textwidth]{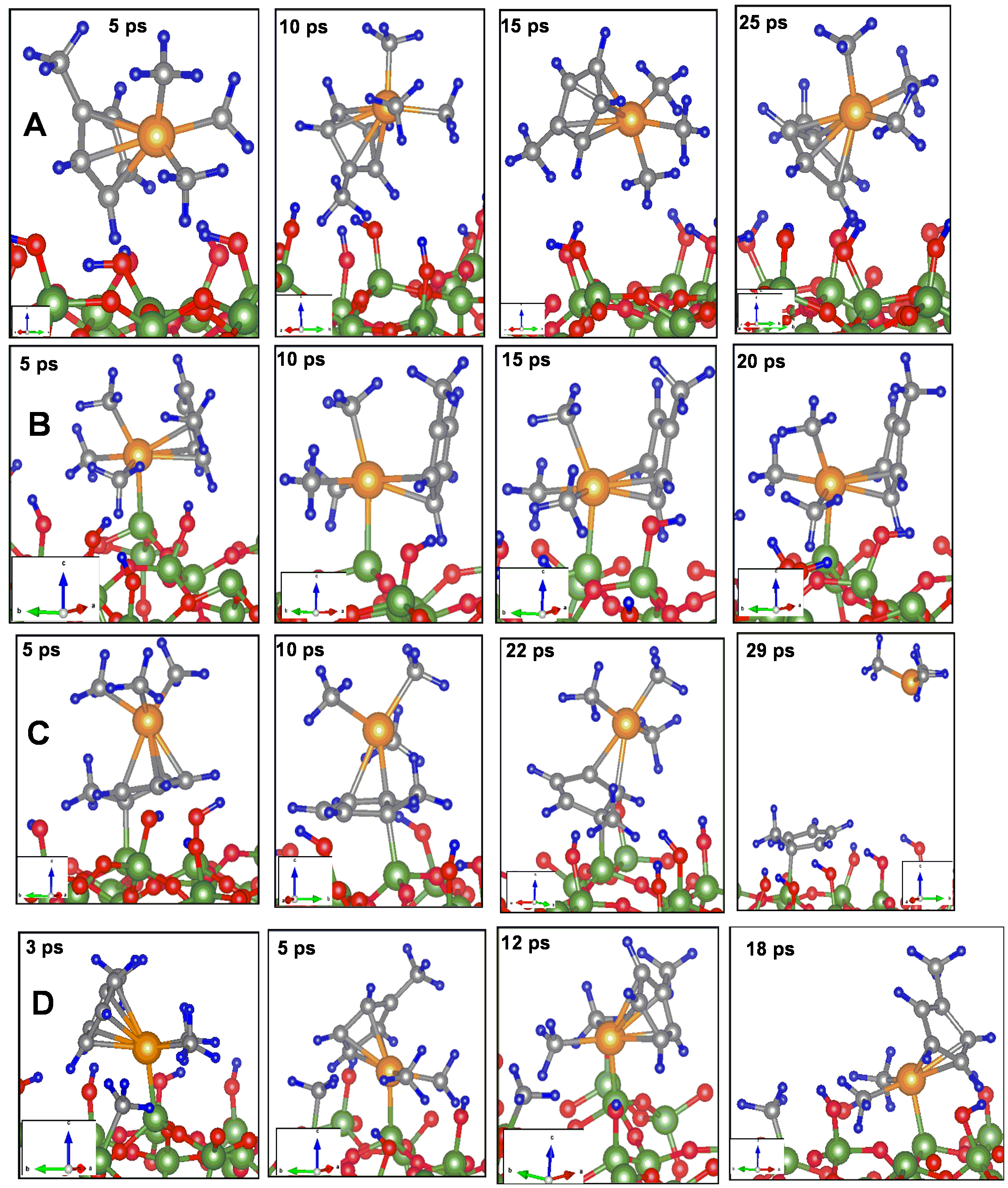}
	\caption{The configurations of {\Pt} obtained from molecular
		dynamics simulations (20 ps) on (a) fully hydroxylated
		surfaces, (b)-(c) 25\% and (d) 50\% partially dehydroxylated
		surfaces}
	\label{fig:lowmdorientation}
\end{figure}

Selected configurations on fully and partially hydroxylated
{\sio} surfaces are computed on this reduced supercell
and compared for consistency ( Table~\ref{tab:secondtable}).
We observe a similar trend for the energetics  and 
the order of stabilization of configurations as in the $3\times 3\times 4$ cell.
For example, the most stable configuration for {\Pt} on the fully
hydroxylated surfaces (Figure~\ref{fig:2x2cellandorientation}~(b))
of both of these cells are similar. The
{\Pt} adsorption on $3\times 3\times 4$ fully hydroxylated surface slabs has
the most stable configuration as OMe$_{2}$-MeCp ($E_{\rm A}=-0.650$
eV) followed by OMe$_{2}$-Cp ($E_{\rm A}=-0.596$ eV)~\cite{Juan2012}.
On a $2\times 2\times 2$ supercell, the most stable configuration is
OMe$_{2}$-MeCp ($E_{\rm A}=-1.02$ eV) followed by OMe$_{2}$-Cp
($E_{\rm A}=-0.845$ eV).  
Analogously, on a $3\times 3\times 4$ partially
hydroxylated surface, the most stable configuration is Model-4
($E_{\rm A}=-1.458$ eV) followed by Model-2 ($E_{\rm A}=-1.443$ eV)
and a similar trend is observed with the reduced supercell i.e., Model-4
($E_{\rm A}=-1.647$ eV) followed by Model-2 ($E_{\rm A}=-1.494$ eV).
Neverthless, the calculated $E_{\rm A}$ on the reduced
supercell is higher than those when using the $3\times 3\times 4$ slabs,
owing to the enhanced concentration of the hydroxyl groups and defective sites
and closer packing of the molecules. 
On the 50\% dehydroxylated cells, where two surface active sites are
located adjacent to each other, one of the {\methyl} groups which was
originally bonded to Pt, fragments and bonds to the adjacent vacant
site, as observed in the $3\times 3\times 4$ cell
(See Fig.~\ref{fig:pohorientation}~(d) and
Fig.~\ref{fig:2x2cellandorientation}~(e)).  These DFT relaxed
structures (as shown in Fig.~\ref{fig:2x2cellandorientation}~(b-e) are
considered as starting point for molecular dynamics simulations.

The trajectory of the MD simulations for the four considered cases are
shown in Fig.~\ref{fig:lowmdorientation}.  The analysis of the
trajectory indicates that on the fully hydroxylated surfaces (
Fig.~\ref{fig:lowmdorientation}~(a)) the molecule exhibits significant
changes in orientation and a drift similar as we found for carbonyl
precursors~\cite{Muthukumar2014}.  The Pt-{Cp} ring and
Pt-{\methyl} fluctuate to a maximum of 5\% and 1\%, respectively. 
Apart from this, in the simulation window, we do not observe any 
further indications on the dissociation of the precursor
molecules.  The trajectory of MD simulations and arising
configurations of {\Pt} on the (25\% and 50\%) partially
dehydroxylated {\sio} surfaces in MD simulations are shown in
Figs.~\ref{fig:lowmdorientation}~(b)-(d).
When simulation is started with the configuration shown in 
Figure~\ref{fig:2x2cellandorientation}~(b), a significant bond weakening
(Pt-{Cp} ring and Pt-{\methyl}) is observed (Fig.~\ref{fig:lowmdorientation}~(b)). Quantitatively, the respective
bond fluctuations computed for Pt-{Cp} ring and Pt-{\methyl} bonds
were 11\% and 9\%, respectively.  However, when MD simulations are performed with
Fig.~\ref{fig:2x2cellandorientation}~(d) as starting configuration,
where a part of the {\cp} ring is bonded to the surface Si atom during
the simulations, the bonding of that part with the Pt(CH$_{3}$)$_{3}$
part of the precursor is weakened.  When simulations are extended further,
the Pt(CH$_{3}$)$_{3}$ part detaches and moves to the vacuum leaving
the {\cp} ring bonded to the substrate (see
Fig.~\ref{fig:lowmdorientation}~(c)).  On the 50\% partially
dehydroxylated surface, (Fig.~\ref{fig:lowmdorientation}~(d)), no
further association or dissociation of methyl groups or detachment of
{\cp} rings are observed.

These results illustrate that the {\Pt} precursor molecules exhibit a
tendency to fragment upon their interaction with the partially
hydroxylated sites and a drifting without fragmentation of molecules
on the fully hydroxylated surfaces.  On the partially hydroxylated surfaces,
the dissociation begins either
with the release of one of the {\Me} groups bonded directly to Pt or
the detachment of the {\cp} ring.  This indicates that surface active
sites are necessary for the fragmentation of precursor molecules and
this is because of the electron density located on the Si atoms, which
is crucial for bonding with the precursor molecule and further
fragmentation~\cite{Muthukumar2011,Muthukumar2012}.  The 25 and 50\%
partially dehydroxylated surfaces represent electron beam pretreated
surfaces and therefore the pretreatment of the substrate with the
electron beam helps in favoring dissociation of the precursor molecule
and efficient deposition.  Since during the MD simulations we do not
observe the dissociation of {\cp} ring or the release of {\methyl}
from the surface Si sites, it is expected that they might block the
active sites from the approach of incoming precursor molecules.

\subsection{Pathways for \Pt~ precursor fragmentation} 

In order to further understand the nature of interaction of {\Pt} with {\POH}
surfaces, the barriers for adsorption and fragmentation of these
molecules to the partially hydroxylated groups were simulated. For
this purpose, a representative simple cluster model Si(OH)$_3$ is
considered. Although this does not account for the complete surface
model, it is a good approximation to evaluate the reaction barriers for  different reactions observed through the MD
simulations.  In the literature, such a simple hydroxylated M(OH)$_{x}$
model has been used as a representative model to investigate
deposition reactions on Al~\cite{Widjaja}, Hf ~\cite{mastail_mechanism} and
Si~\cite{Hu,ABM,ADH}.  Also, the energies reported in this section are
computed at the PM6 level of theory, and a relative comparison should
enable reasonable understanding of the possible fragmentation pathways
qualitatively. In a recent investigation, thickness controlled site-selective
Pt deposits were obtained by the direct ALD (Atomic layer deposition) 
process, in which ALD was performed on EBID patterned substrates. 
In this ALD process, an {\otwo} pulse is used to obtain better nucleation, even though the qualitative understanding of the role of {\otwo} remains uncertain. 
Therefore, in this section we consider  involvement of {\otwo} in the pathways
at different stages of the reaction and calculate the energetics to compare and elucidate the role of \otwo \cite{ald+ebid1}.

\begin{figure*}
	\centering
	\includegraphics[width=1\textwidth]{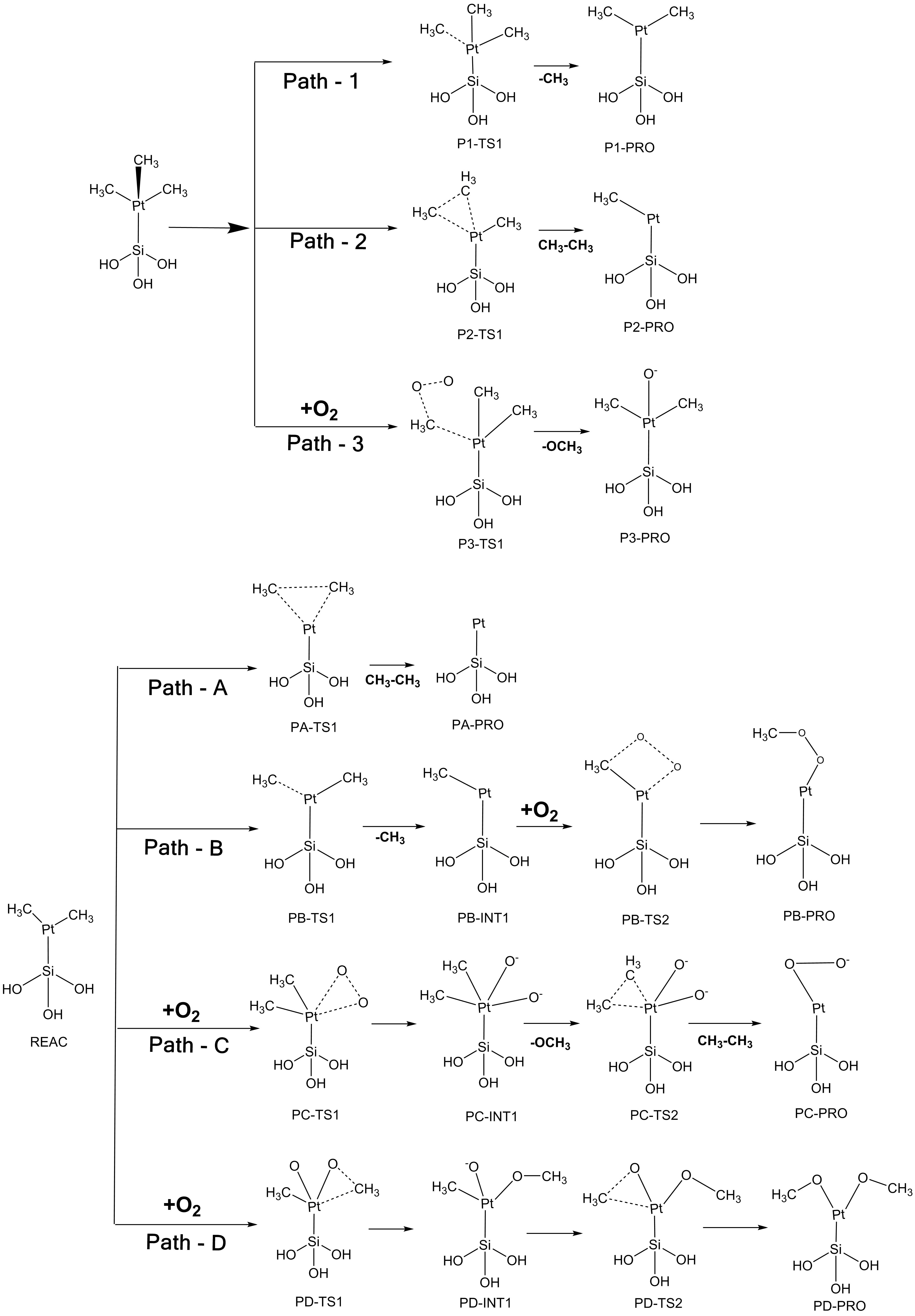}
	\caption{Possible reaction channels for the dissociation of
		{\Pt} after its interaction with {\POH} surfaces and in an
		{\otwo} environment. Possible pathways for the (a) release of
		a first methyl group and (b) subsequent methyl groups.}
	\label{fig:scheme1}
\end{figure*}

\begin{figure*}
	\centering
	\includegraphics[width=0.98\textwidth]{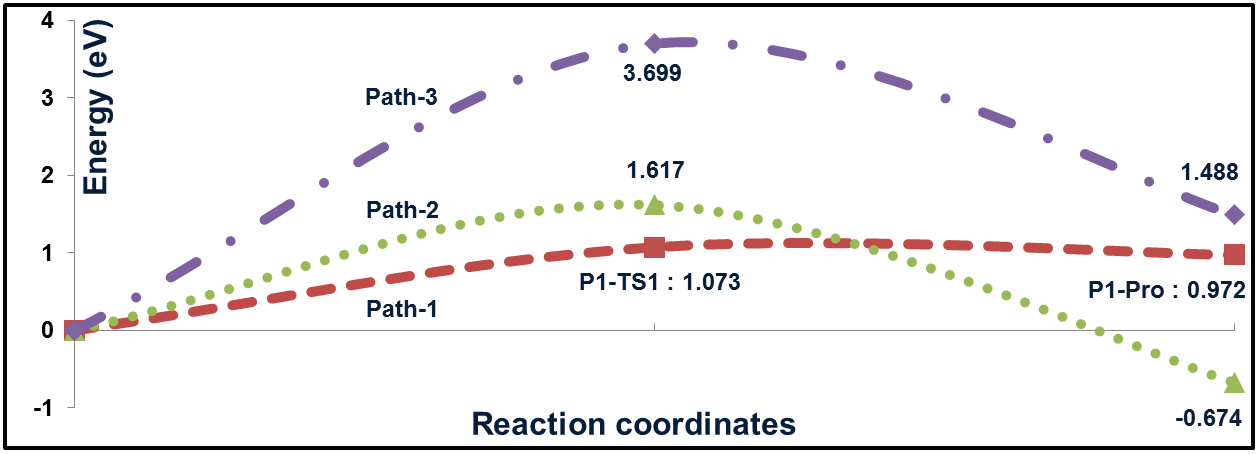}
	\includegraphics[width=1\textwidth]{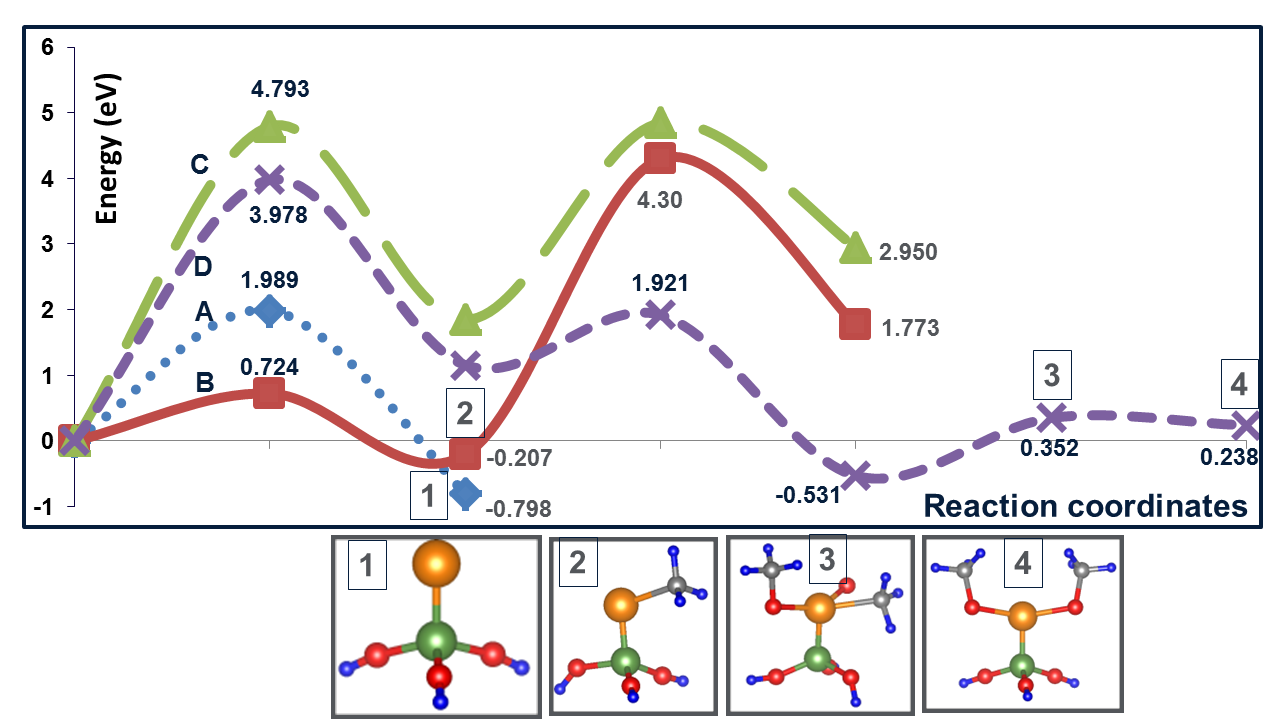}
		\caption{Energetics for different pathways (PATH 1-3 and
		A-D) considered for the dissociation of precursor
		molecules. Representative figures are displayed.}
	\label{fig:scheme3}
\end{figure*}

Our simulations of the adsorption and dynamics of {\Pt} on \sio\ surfaces
show that the dissociation begins with either the detachment of 
{\methyl} groups or the {\cp} ring. 
In the cluster model, the  interaction of {\Pt} 
with the surface Si atoms, has a barrier of 
0.140~eV, when {\Pt} interacts through the {\cp} ring.  
Removal of the {\cp} ring from {\Pt} has an activation energy
of +0.771~eV, which leaves the Pt(CH$_{3}$)$_{3}$ part of the
precursor binding to the substrate. 
Formation of Pt(CH$_{3}$)$_{3}$ on vacuum is observed in our molecular dynamics 
simulations. However, neither its dissociation into vacuum or 
adsorption back to the surface Si sites has not been observed.
Therefore, for analyzing the dissociation channels of
Pt(CH$_{3}$)$_{3}$, we reasonably approximate 
our starting configuration as Pt(CH$_{3}$)$_{3}$ bonded with the 
surface Si site as shown in Figure~\ref{fig:scheme1}. 
Our adsorption studies reveal that for the elimination of
methyl groups, two surface active sites are necessary, and the present
model limits the computation of barrier for such cases.   

Possible pathways by which Pt(CH$_{3}$)$_{3}$ gets dissociated on the model cluster, in the presence and absence of
{\otwo}, are derived and shown in Scheme~\ref{fig:scheme1}.  From Pt(CH$_{3}$)$_{3}$, the release of one {\methyl} group (P1TS1) forming Pt(CH$_{3}$)$_{2}$ has an activation
energy of about +1.073~eV compared to +1.617~eV for ethane release
(P2TS1). The {\otwo} assisted elimination of first methyl (P3TS1) from
Pt(CH$_{3}$)$_{3}$ has a barrier of +3.699~eV. Therefore, release of a
first {\methyl} is expected to proceed through P1TS1, leaving
Pt(CH$_{3}$)$_{2}$ (P1-PRO-1) on the Si surface.
The removal of the second and third {\methyl} from Pt(CH$_{3}$)$_{2}$
can happen by a number of ways ((PATH A-D, see Fig.~\ref{fig:scheme1}
and Fig.~\ref{fig:scheme3}).  
Ethane elimination, which leads to a Pt atom (PA-TS1) bonded to the Si
surface has a barrier of about +1.989~eV.  This reaction is exothermic
by -0.798~eV.  The first step (PBTS-1) in Path-B, leading to second
methyl elimination has a barrier of +0.724~eV, and occurs in a
similar fashion as the first methyl removal. This reaction resulting in species
(PBINT-1) is exothermic by -0.207~eV.  Comparing the pathways
considered and the computed activation energies as shown in
Figure~\ref{fig:scheme3}), the third methyl elimination in the presence
of oxygen and the subsequent formation of Pt-oxy species (Path-B, C, D)
have a high activation barrier.  

The formed Pt or Pt-oxy species might agglomerate and form Pt deposits
and the role of {\otwo} on this process hasn't been
explored in this investigation. 
Some experimental studies were available and possible
mechanisms were proposed~\cite{Stanford}.  Simulations in this
study were performed at the PM6 level, with no corrections for weak
interactions and at $T=0$~K.  Taking into account the temperature of
most processess, which range between $200 and 400\,^{\circ}{\rm C}$, some of the high-barrier pathways might also be operative. Also, the
role of secondary electrons and the point defects which might occur as
a result of electron impingement on the surface have not been explored. This study therefore provides a preliminary
understanding of how the {\Pt} molecule can dissociate on the
substrates that qualitatively represent electron beam pretreated
surfaces.

\section{CONCLUSIONS}\label{Summarize}

We have performed theoretical simulations on the dynamics and
fragmentation mechanisms of {\Pt} on the fully and partially
hydroxylated {\sio} surfaces.  Our results provide  clues
for the most stable
configurations of {\Pt} on {\sio} surfaces and their dynamical
behavior. These results illustrate that the fragmentation of the
precursor molecule begins with either the detachment of either the {\cp}
ring or  the methyl group. Detached {\cp} rings and the
dissociated {\methyl} groups block the surface active sites and might
be the source of organic contamination.  Since the composition of the
obtained deposits dictate the conductance behavior, it can be
speculated that a design of suitable precursors for electron beam
induced deposition might be more efficient
 than the use of traditional ALD precursors. 
 With our reaction modeling studies, possible pathways
by which the precursor molecule can fragment on {\sio} surfaces were
also explored.

\section{Acknowledgments}\label{Acknowledgments}
The authors gratefully acknowledge financial support by the
Beilstein-Institut, Frankfurt/Main, Germany, and the EU COST action CELINA.
  The generous allotment of computer time by
CSC-Frankfurt and LOEWE-CSC is gratefully acknowledged.


\begin{thebibliography}{10}


\bibitem{Wnuk2011}
Wnuk, J. D.; Rosenberg, S. G.; Gorham, J. M.; van Dorp, W. F.; Hagen, C. W.; Fairbrother, D. H. 
{\it Surf. Sci.} {\bf 2011}, {\it 605}, 257.

\bibitem{Utke2008}
Utke, I.; Hoffmann, P.; Melngailis, J. 
{\it J. Vac. Sci. Technol. B} {\bf 2008}, {\it 26}, 1197.

\bibitem{weber:461}
Weber, M.; Rudolph, M.; Kretz, J.; Koops, H. W. P. 
{\it J. Vac. Sci. Technol. B} {\bf 1995}, {\it 13}, 461.

\bibitem{Wnuk2009}
Wnuk, J. D.; Gorham, J. M.; Rosenberg, S. G.; van Dorp, W. F.; Madey, T. E.; Hagen, C. W.; Fairbrother, D. H.
{\it J. Phys. Chem. C} {\bf 2009}, {\it 113}, 2487.

\bibitem{Weirich2013}
Weirich, P. M.; Winhold, M.; Schwalb, C. H.; Huth, M.
{\it Beilstein J. Nanotechnol.} {\bf 2013}, {\it 4}, 919.

\bibitem{incompletedissociation1}
Hedhili, M. N.;  Bredehft, J. H.; Swiderek, P. 
{\it J. Phys. Chem. C} {\bf 2009}, {\it 113}, 13282.

\bibitem{Huthreview2012}
Huth, M.; Porrati, F.; Schwalb, C.; Winhold, M.; Sachser, R.; Dukic, M.; Adams, J.; Fantner, G.
{\it Beilstein J. Nanotechnol.} {\bf 2009}, {\it 3}, 597.

\bibitem{Lewis2015}
 Lewis, B. B.; Stanford, M. G.; Fowlkes, J. D.; Lester, K.; Plank, H.; Rack, P. D. 
{\it Beilstein J. Nanotechnol.} {\bf 2015}, {\it 6}, 907.

\bibitem{Muthukumar2011}
Muthukumar, K.; Opahle, I.;  Shen, J.; Jeschke, H. O.; Valent{\'\i}, R.
{\it Phys. Rev. B} {\bf 2011}, {\it 84}, 205442.

\bibitem{Muthukumar2012}
Muthukumar, K.; Jeschke,  H. O.; Valent{\'\i}, R.; Begun, E.; Schwenk, J.; Porrati, F.; Huth,  M.
{\it Beilstein J. Nanotechnol.} {\bf 2012}, {\it 3}, 546.

\bibitem{Muthukumar2014}
Muthukumar, K.; Jeschke, H. O.; Valent{\'\i}, R. 
{\it J. Chem. Phys.} {\bf 2014}, {\it 140}, 184706.

\bibitem{ald-smgeorge}
George,  S. M. 
{\it Chem. Rev.} {\bf 2010}, {\it 110}, 111.

\bibitem{mastail_mechanism}
Mastail, C.; Lanthony, C.; Olivier, S.; Duc{\'e}r{\'e}, J.-M.; Landa, G.; Est{\`e}ve, A.; Rouhani, M. D.; Richard, N.; Dkhissi, A.
{\it Thin Solid Films} {\bf 2012}, {\it 520}, 4559.

\bibitem{Juan2012}
Shen, J.; Muthukumar, K.; Jeschke, H. O.; Valent{\'\i}, R.
{\it New. J. Phys.} {\bf 2012}, {\it 14}, 073040.

\bibitem{Bloechl1994}
Bl\"ochl, P. E. 
{\it Phys. Rev. B} {\bf 1994}, {\it 50}, 17953.

\bibitem{Kresse1999}
Kresse, G.; Joubert, D.
{\it Phys. Rev. B} {\bf 1999}, {\it 59}, 1758.

\bibitem{Kresse1996a}
Kresse, G.; Furthm\"uller, J. 
{\it Phys. Rev. B} {\bf 1996}, {\it 54}, 11169.

\bibitem{Kresse1996b}
Kresse, G.; Furthm\"uller, J. 
{\it Comput. Mater. Sci.} {\bf 1996}, {\it 6}, 15.

\bibitem{Kresse1993}
Kresse, G.; Hafner, J.
{\it Phys. Rev. B} {\bf 1993}, {\it 47}, 558.

\bibitem{Perdew1996} 
Perdew, J. P.; Burke, K.; Ernzerhof, M.
{\it Phys. Rev. Lett.} {\bf 1996}, {\it 77}, 3865.

\bibitem{Grimme2006}
Grimme, S. 
{\it J. Comput. Chem.} {\bf 2006}, {\it 27}, 1787.

\bibitem{JCC:JCC21057}
Hafner, J. 
{\it J. Comput. Chem.} {\bf 2008}, {\it 29}, 2044.

\bibitem{temperature}
Dendooven, J.; Ramachandran, R. K.; Casier, K. D.; Rampelberg, G.; Filez, M.; Poelman, H.; Marin, G. B.; Fonda, E.; Detavernier, C.
{\it J. Phys. Chem. C} {\bf 2013}, {\it 117}, 20557.


\bibitem{go9}
Gaussian 09, Revision A.02; Gaussian, Inc.: Wallingford, CT, 2016.

\bibitem{Egger}
Egger, K. W. 
{\it J. Organometal. Chem.} {\bf 1970}, {\it 24}, 501.

\bibitem{Pt2014}
Spencer J. A.; Rosenberg, G. S.; Barclay, M.; Wu Y-C.; McElwee-White L.; Fairbrother H.D.;
{\it Appl. Phys. A} {\bf 2014}, {\it 117}, 1631.

\bibitem{Liang}
Liang, X.; Zhou, Y.; Li, J.; Weimer, A. W.
{\it J. Nanopart. Res.} {\bf 2011}, {\it 13}, 3781.

\bibitem{Koplitz}
Koplitz, L. V.; Shuh, D. K.; Chen, Y.-J.; Williams, R. S.; Zink, J. I.
{\it Appl. Phys. Lett.} {\bf 1988}, {\it 53}, 1705.

\bibitem{Widjaja}
Widjaja, Y.; Musgrave, C. B.
{\it Appl. Phys. Lett.} {\bf 1999}, {\it 80}, 3304.

\bibitem{ABM}
Mukhopadhyay, A. B.; Musgrave, C. B.
{\it Chem. Phys. Lett.} {\bf 2006}, {\it 421}, 215.

\bibitem{Hu}
Hu, Z.; Turner, C. H.
{\it J. Phys. Chem. B} {\bf 2006}, {\it 110}, 8337.

\bibitem{ADH}
Dkhissi, A.; Esteve, A.; Jeloaica, L.; Djafari Rouhani, M.; Landa, G.
{\it Chem. Phys.,} {\bf 2006}, {\it 323}, 179.

\bibitem{ald+ebid1}
Mackus, A. J. M.; Thissen, N. F. W.; Mulders, J. J. L.; Trompenaars, P. H. F.; Verheijen, M. A.; Bol, A. A.; Kessels,  W. M. M. 
{\it J. Phys. Chem. C} {\bf 2013}, {\it 117}, 10788.

\bibitem{Stanford}
Stanford, M. G.; Lewis, B. B.; Noh J. H.; Fowlkes, D. J.; Roberts, A. N.; Plank, H; Rack, P.D;
{\it ACS Appl. Mater. Interfaces} {\bf 2014}, {\it 6}, 21256.

\end{thebibliography}
\end{document}